\documentclass[reprint,10pt,twocolumn,superscriptaddress,nofootinbib]{revtex4-1}

\usepackage[utf8]{inputenc}
\usepackage{graphicx}
\usepackage{bm}
\usepackage[colorlinks=true]{hyperref} 

\hypersetup{
    colorlinks=true,
    linkcolor=blue,
    filecolor=magenta,      
    urlcolor=cyan,
}
\usepackage{amsmath,amssymb,amsbsy}
\usepackage{bbm,mathrsfs,yfonts,eufrak}
\usepackage[usenames,dvipsnames]{color}

\newcommand{\avgF}[1]{\ensuremath{\left[ \left[#1 \right]\right]}}
\newcommand{\flucF}[1]{\ensuremath{\widehat{ #1  }}}

\newcommand{\avgR}[1]{\ensuremath{\langle #1 \rangle}}
\newcommand{\flucRrel}[1]{\ensuremath{\delta{ #1} }}
\newcommand{\flucR}[1]{\ensuremath{\widetilde{ #1 }}}
\renewcommand{\ne}{\ensuremath{n }}
\newcommand{\neref}{\ensuremath{n_0}}
\newcommand{\Ni}{\ensuremath{N}}

\newcommand{\OmegaciN}{\ensuremath{\Omega_0}} 
\newcommand{\rhoN}{\ensuremath{\rho_{s0}}} 
\newcommand{\csN}{\ensuremath{c_{s0}}} 
\newcommand{\teN}{\ensuremath{T_{e0}}} 

\newcommand{\vordf}{\ensuremath{\mathcal{W}_\delta}} 
\newcommand{\vorfF}{\ensuremath{\mathcal{W}}} 
\renewcommand{\vec}[1]{{\mathbf{#1}}}
\newcommand{\bhat}{\hat{\vec b}}

\begin{document}
\title{Non-Oberbeck-Boussinesq zonal flow generation}
\author{M.\ Held}
\email[E-mail: ]{markus.held@uibk.ac.at}
\affiliation{Institute for Ion Physics and Applied Physics, 
                     Universit\"at Innsbruck, A-6020 Innsbruck, Austria}
\author{M.\ Wiesenberger}
\affiliation{Department of Physics, Technical University of Denmark, DK-2800 Kgs. Lyngby, Denmark}
\author{R.\ Kube}
\affiliation{Department of Physics and Technology, UiT The Arctic University of Norway, N-9037 Troms{\o}, Norway}
\author{A.\ Kendl}
\affiliation{Institute for Ion Physics and Applied Physics, 
                     Universit\"at Innsbruck, A-6020 Innsbruck, Austria}
\date{\today}
\begin{abstract} 
Novel mechanisms for zonal flow (ZF) generation for both large relative density fluctuations and background density gradients are presented.
In this non-Oberbeck-Boussinesq (NOB) regime ZFs are driven by the Favre stress, the large fluctuation extension of the Reynolds stress, 
and by background density gradient and radial particle flux dominated terms. 
Simulations of a nonlinear full-F gyro-fluid model confirm the predicted mechanism for radial ZF propagation and show 
the significance of the NOB ZF terms for either large relative density fluctuation levels or steep background density gradients.
\end{abstract}
\maketitle 
\section{Introduction}
Self-organization from turbulent to coherent states is a ubiquitous process in fluids.
In particular, much interest and effort has been drawn to the formation of zonal flows (ZFs)~\cite{diamond05,fujisawa09,guercan15}.
These coherent flows arise in 
atmospheres, in the form of banded cloud structures on Jupiter~\cite{heimpel05}, Saturn's north-polar hexagon~\cite{baines09} 
or mid-latitude westerlies on earth and in the ocean as stationary jets~\cite{maximenko05}.
In magnetized fusion plasmas ZFs are key players for the reduction of the radial transport of particles and heat 
and for the transition to improved confinement regimes in tokamaks~\cite{burrell97,terry00,hillesheim16,tynan16,schmitz17,cziegler17}.

Reynolds stress is quintessential for ZF generation in all fluids~\cite{diamond91,xu00,diamond05,fujisawa09,mueller11,xu11,yan14,guercan15},
but in magnetized plasmas also other stresses like the Maxwell~\cite{craddock91,scott05} or the diamagnetic stress~\cite{smolyakov00,madsen17} can become significant.
Virtually all of the work on ZF theory so far rely on \(\flucRrel{f}\) models~\cite{diamond91,diamond05,guo16}, 
which invoke the so called Oberbeck-Boussinesq (or thin layer) approximation~\cite{oberbeck1879,boussinesq03}.
However, the latter breaks down, if the background density varies over more than one order of magnitude or 
if the relative density fluctuations exceed roughly \(10\) percent. 
This for example prevails in the edge of tokamak fusion plasmas, where experimental measurements typically feature relative density fluctuation levels around the order \(0.1\) in the edge and up 
to unity at the last closed flux surface~\cite{surko83,liewer85,ritz87,fonck93,endler95,mckee01,zweben07,zweben15,gao15}.
Moreover, typical edge background density gradient (e-folding) lengths reach from \(50 \rhoN\) in low-confinement to  \(10 \rhoN\) in high-confinement tokamak plasmas~\cite{shao16,kobayashi16}. Here, 
\(\rhoN:=\sqrt{\teN m_i }/(e B_0)\) is the drift scale with reference electron temperature \(\teN\), ion particle charge \(e\), ion mass \(m_i\) and reference magnetic field \(B_0\). 

Non-Oberbeck-Boussinesq (NOB) effects on ZF generation are an unresolved issue. However, 
theoretical and experimental studies of poloidal ZFs in the edge of fusion plasmas indicate that unknown
mechanisms beyond the Reynolds stress exist~\cite{kobayashi13} and that steep background density gradients and large relative density fluctuations affect the poloidal ZF dynamics~\cite{mueller11,windisch11,wang16}.
Moreover, the importance of large relative density fluctuations for toroidal momentum transport, as suggested by theoretical estimates in the strong and weak turbulence regime~\cite{wang15,kosuga17} and experimental measurements in the TORPEX and PANTA device~\cite{labit11,inagaki16}, point towards a similar significance for poloidal momentum transport.

In the following we generalize the theory of ZFs to NOB effects.
To this end, we decompose the density and electric potential of a full-F gyro-fluid model of a magnetized plasma~\cite{madsen13} with the help of a density weighted Favre average~\cite{favre65}.
This  well known decomposition strategy in compressible fluid dynamics (see e.g.~\cite{wilcox00}) is here for the first time introduced to plasma physics and enables us to 
disentangle the density fluctuations from the ZF dynamics, while retaining the relevant physical effects. 
As a result, we identify novel agents in the poloidal ZF dynamics, which become significant 
for high relative density fluctuations or steep background density gradients.
We confirm the herein proposed NOB mechanism for radial advection of ZFs with the help of numerical simulations 
of a fully nonlinear model for drift wave-ZF dynamics. 
The exploited model is based on the specified extension of the Hasegawa-Wakatani model to the full-F framework. 
Additionally, we show how the ZF dynamics is distributed among the proposed NOB actors and provide 
scalings with collisionality, reference background density gradient length and the maximum of the relative density fluctuation amplitude. 
\section{ZF theory}
\subsection{\(\flucRrel{f}\) formalism}
We start our discussion with a short re-derivation of the conventional ZF equation and Reynolds stress from a cold ion 
\(\flucRrel{f}\) gyro-fluid model, which couples small relative density fluctuations to the electric potential via \(\vec{E}\times\vec{B}\) advection and linear polarization~\cite{dorland93,beer96,scott10a}
\begin{subequations} 
\label{eq:dfmodel}
\begin{eqnarray}
\label{eq:deltafdtne}
 \frac{\partial}{\partial t }\flucRrel{\ne} + \vec{\nabla} \cdot \left(\flucRrel{\ne} \vec{u}_E \right) +\frac{1}{L_n B_0}  \frac{\partial }{\partial y} \phi  = \Lambda_\delta
,\\
\label{eq:deltafdtNi}
 \frac{\partial}{\partial t }\flucRrel{\Ni} + \vec{\nabla} \cdot \left( \flucRrel{\Ni} \vec{u}_E \right) +\frac{1}{L_n B_0}  \frac{\partial }{\partial y}\phi  = 0,\\
\label{eq:dfpolcold}
 \vec{\nabla} \cdot \left(\frac{1}{\OmegaciN} \frac{\vec{\nabla}_\perp \phi}{B_0}\right)= \flucRrel{\ne} - \flucRrel{\Ni}.
\end{eqnarray}
\end{subequations}
Here, \(\flucRrel{\ne}:= \ne/n_G-1\) is the relative electron density fluctuation, \(\flucRrel{\Ni}:= \Ni/n_G-1\) is the relative ion gyro-center density fluctuation, \(\phi\) is the electric potential and 
\(\OmegaciN :=  e B_0/m_i\) is the ion gyro-frequency. 
The reference background density \(n_G(x)\) refers to a constant reference background gradient length \(L_n:= -1/\partial_x \ln{(n_G/\neref)} \) with constant reference density \(\neref\).
For the sake of simplicity the magnetic field \(B=B_0\) is assumed constant and the unit vector in the magnetic field direction is \(\bhat := \vec{B}/B_0 = \vec{\hat{e}}_z\). 
The perpendicular gradient and the \(\vec{E}\times\vec{B}\) drift velocity are defined by \(\vec{\nabla}_\perp := - \bhat \times (\bhat \times \vec{\nabla})\) and  
\(\vec{u}_E:= \bhat \times \vec{\nabla} \phi/B_0\), respectively. 
The term \(\Lambda_\delta\) denotes a closure for the parallel dynamics, which is discussed later in more detail.
Taking the time derivative over the polarization equation~\eqref{eq:dfpolcold} yields the \(\flucRrel{f}\) drift-fluid vorticity density equation
\begin{eqnarray}\label{eq:dtvordf}
    \frac{\partial}{\partial t } \vordf  + \vec{\nabla} \cdot \left(\vordf\vec{u}_E \right) = \OmegaciN \neref \Lambda_\delta,
\end{eqnarray}
with the linear \(\vec{E}\times\vec{B}\) vorticity density \(\vordf :=\neref\vec{\nabla}_\perp^2 \phi/B_0 =  \bhat \cdot \vec{\nabla} \times \left( \neref \vec{u}_E\right)\).
Now we apply the average over the ``poloidal'' y coordinate \(\avgR{  h} := L_y^{-1}\int_0^{L_y} dy \hspace{1mm} h\) to Eq.~\eqref{eq:dtvordf},
which is the 2D equivalent  of a flux surface average. 
Reynolds decomposition \(h = \avgR{h} + \flucR{h}\) and integration over the ``radial'' coordinate \(x\) result in 
the \(\flucRrel{f}\) evolution equation for poloidal ZFs~\cite{diamond91}
\begin{eqnarray}\label{eq:deltafmeanflow}
\frac{\partial }{\partial t}  \avgR{   u_{y} }  =& - \frac{\partial}{\partial x}  \avgR{ \flucR{u_{x} } \flucR{u_{y}} }
                                                       +  \OmegaciN\int_{x_0}^x d x \avgR{ \Lambda_\delta }  .              
\end{eqnarray}
Here, we introduced \(u_{x} := - \partial_y \phi/B_0\), \(u_{y} := \partial_x \phi/B_0\) and 
the anticipated Reynolds stress  \(\mathcal{R} := \avgR{  \flucR{u_{x}} \flucR{u_{y}} } \)~\cite{reynolds1895}, where \(\avgR{  u_{x} u_{y} } = \avgR{  u_{x} }  \avgR{  u_{y} } + \avgR{  \flucR{u_{x}} \flucR{u_{y}} }\)
and \( \avgR{  u_{x} } = 0\) was used. 
In passing we note that we assume that radial boundary conditions give rise to no additional terms in Eq.~\eqref{eq:deltafmeanflow} and for the remainder of this letter.
\subsection{Full-F formalism}
In full-F theory the splitting of the gyro-fluid moment variables into fluctuating and background parts is avoided and 
the quasi-neutrality constraint for electrons and ions is rendered by the nonlinear polarization equation~\cite{madsen13}.
The cold ion full-F gyro-fluid model~\cite{Wiesenberger2014,held16b}
\begin{subequations} 
\label{eq:ffmodel}
\begin{eqnarray}
\label{eq:fullFdtne}
 \frac{\partial}{\partial t }\ne  +\vec{\nabla} \cdot \left( \ne \vec{u}_E\right) =\Lambda
,\\
\label{eq:fullFdtNi}
\frac{\partial}{\partial t } \Ni +\vec{\nabla} \cdot \left( \Ni \vec{U}_E\right) =0 , \\
\label{eq:ffpolcoldlwl}
 \vec{\nabla} \cdot \left(\frac{\Ni}{\OmegaciN} \frac{\vec{\nabla}_\perp \phi}{B_0}\right)= \ne - \Ni, 
\end{eqnarray}
\end{subequations}
evolves the full electron density \({\ne}\) and ion gyro-center density \(\Ni\). 
In the gyro-center \(\vec{E}\times\vec{B}\) drift velocity  by 
\(\vec{U}_E:= \vec{u}_E + \vec{U}_p\) the ponderomotive correction \( \vec{U}_p :=-\bhat \times \vec{\nabla} \vec{u}_E^2/(2 \OmegaciN)\) appears. 
Both, the latter ponderomotive correction and the polarization charge nonlinearity on the left hand side of Eq.~\eqref{eq:ffpolcoldlwl} are crucial for energetic 
consistency and an exact momentum conservation law~\cite{scott10b}.
We refer to the parallel closure term \(\Lambda\) 
later on.
In the long wavelength limit we can again reformulate Eqs.~\eqref{eq:fullFdtNi} and~\eqref{eq:ffpolcoldlwl} into a drift-fluid vorticity density equation
\begin{eqnarray} \label{eq:dtvorfullF}
  \frac{\partial}{\partial t } \vorfF  + \vec{\nabla} \cdot \left(\vorfF\vec{u}_E \right)  - \OmegaciN \vec{\nabla} \cdot \left(n \vec{U}_p \right) =\OmegaciN \Lambda,
\end{eqnarray}
where the nonlinear \(\vec{E}\times\vec{B}\) vorticity density is given by \(\vorfF := \vec{\nabla} \cdot \left(\ne \vec{\nabla}_\perp \phi/B_0\right) 
=  \bhat \cdot \vec{\nabla} \times \left( n \vec{u}_E\right) \).
As above, we obtain the averaged poloidal momentum equation ~\cite{guercan07,shurygin12,wang16}
\begin{align}\label{eq:fullfmeanflow}
\frac{\partial }{\partial t} \avgR{  \ne u_{y} }=& 
-\frac{\partial}{\partial x} \left(
                                        \avgR{  \ne }   \mathcal{R}
                                       +\avgR{  \flucR{\ne}  \flucR{u_{x}}} \avgR{  u_{y} }
                                     +\avgR{  \flucR{\ne}  \flucR{u_{x}}\flucR{u_{y}} } \right)
                                     \nonumber  \\ 
                                     &+\OmegaciN\int_{x_0}^x d x \avgR{  \Lambda } .
\end{align}
The divergence of the full-F stress drive terms of Eq.~\eqref{eq:fullfmeanflow} is related to the averaged radial 
flux of vorticity density minus the ponderomotive correction via the full-F Taylor identity
 \begin{align}\label{eq:taylorfullF}
  \avgR{\flucR{u_x} \flucR{\vorfF}}   =& 
  \frac{\partial}{\partial x} \big[ \avgR{\ne} \mathcal R + \avgR{\flucR{\ne} \flucR{u_x}} \avgR{u_y}
  +  \avgR{\flucR{\ne} \flucR{u_x} \flucR{u_y}}\big] \nonumber \\
  &+\OmegaciN \avgR{\flucR{U_{p,x}} \flucR{n}} .
 \end{align}
The interpretation of Eq.~\eqref{eq:fullfmeanflow} is problematic since
(i) absolute density fluctuations \(\flucR{\ne}\) arise instead of relative density fluctuations  \(\flucR{\ne}/\avgR{\ne}\), 
(ii) the time evolution of the averaged poloidal momentum \(\avgR{  \ne u_{y} }\) is given in terms of the averaged poloidal velocity \(\avgR{  u_{y} }\)
and 
(iii) background density gradient \(\partial_x \ln{\avgR{  \ne }} \) effects are not obvious.
Despite these obstacles Eq.~\eqref{eq:fullfmeanflow} has been recently used to show that 
the second term, occasionally misinterpreted as advective, and the cubic term can be comparable to the 
Reynolds stress related drive \(\partial_x (\avgR{\ne} \mathcal{R}) \)~\cite{mueller11,windisch11,wang16}.

Thus, we go a step further and utilize a density weighted Favre decomposition instead of the Reynolds decomposition according to
\(  h := \avgF{h} + \flucF{h} \) and  \( \avgF{h} :=\avgR{  \ne h}/\avgR{  \ne } \)~\cite{favre65}.
Note that the Favre decomposition reduces to the Reynolds decomposition if the density \(\ne\) is only a function of \(x\). 
Now we combine the poloidal average of Eq.~\eqref{eq:fullFdtne} divided by \(\avgR{  \ne }\)
\begin{align}\label{eq:fullflognemeanflow}
\frac{\partial }{\partial t}\ln{\avgR{  \ne }}  =& - \frac{\partial}{\partial x}  \avgF{ u_{x}} - \avgF{ u_{x}}\frac{\partial}{\partial x} \ln{\avgR{  \ne }}
\nonumber \\&+ \frac{\avgR{  \Lambda }}{\avgR{  \ne }}
\end{align}
with  Eq.~\eqref{eq:fullfmeanflow} divided by \(\avgR{  \ne }\)
to obtain a ZF evolution equation for the Favre averaged poloidal velocity
\begin{align}\label{eq:fullfmeanflowcombined}
\frac{\partial }{\partial t}  \avgF{u_{y}}  =&  - \frac{\partial}{\partial x}  \avgF{\flucF{u_{x}} \flucF{u_{y}} }
						- \avgF{u_{x}} \frac{\partial}{\partial x}  \avgF{ u_{y}}  
\nonumber \\&
                                                       - \avgF{\flucF{u_{x}} \flucF{u_{y}} }  \frac{\partial}{\partial x} \ln{\avgR{  \ne }}      
				    \nonumber \\&
                                                       -  \avgF{u_{y}}  \frac{\avgR{  \Lambda }}{\avgR{  \ne }}
                                                       +  \frac{\OmegaciN}{\avgR{  \ne }}\int_{x_0}^x d x \avgR{ \Lambda } ,
\end{align}
where we used  \(\avgF{ u_{x} u_{y} } = \avgF{u_{x}}    \avgF{u_{y}}+\avgF{  \flucF{u_{x}} \flucF{u_{y}} }\).
The Favre stress  \(\mathcal{F} := \avgF{\flucF{u_{x}}\flucF{u_{y}}}\) can be rewritten into
\begin{eqnarray}\label{eq:Favrestress}
 \mathcal{F}    = \mathcal{R} - \avgF{  \flucR{u_{x}}} \avgF{\flucR{u_{y}}}
                                       +\avgR{  \flucR{\ne}  \flucR{u_{x}}\flucR{u_{y}} } / \avgR{  \ne } .
\end{eqnarray}
Consequently, the first term \(-\partial_x \mathcal{F}\) on the right hand side of Eq.~\eqref{eq:fullfmeanflowcombined} is the superposition of the conventional Reynolds stress drive 
\(T_1 := -\partial_x \mathcal{R}\), the quadruple fluctuation term \(T_2 := \partial_x \left( \avgF{  \flucR{u_{x}}} \avgF{\flucR{u_{y}}} \right)\) and the
triple fluctuation drive \(T_3 := -\partial_x \left( \avgR{  \flucR{\ne}  \flucR{u_{x}}\flucR{u_{y}} } / \avgR{  \ne }  \right)\). 
The novel second term on the right hand side of Eq.~\eqref{eq:fullfmeanflowcombined} represents radial advection of poloidal ZFs \(\avgF{u_{y}}\). 
Its direction depends on the sign of the averaged radial particle flux \(\avgR{  \Gamma_{x}}:= \avgR{  \ne }\avgF{u_{x}}\), which is typically positive, so that \(T_4\) describes an outward pinch of ZFs.
The novel third term \(T_5 := \mathcal{F} /L_{\avgR{  n }}\)  on the right hand side of Eq.~\eqref{eq:fullfmeanflowcombined} is proportional 
to the inverse of the background density gradient length \(1/L_{\avgR{  n }}:= -\partial_x \ln{\avgR{  \ne }}\). 
This term is large for small reference background density gradient lengths \(L_n\), or has large radially localized values if the density profile \(\avgR{n}\) develops into a staircase like pattern~\cite{difpradalier15}.
In contrast to the Favre stress drive \(-\partial_x  \mathcal{F} \),
the background density gradient drive \(T_5\) contributes to the ZF generation even if the Favre stress is 
radially homogeneous \(\partial_x \mathcal{F}=0\). Remarkably, the background density gradient drive remains finite in the small relative density  fluctuation limit, where 
the density \(n\) is only a function of \(x\) and the Favre stress \(\mathcal{F}\) resembles the conventional Reynolds stress \(\mathcal{R}\).

In order to interpret  the dynamics of the background density gradient drive \(T_5\) let us assume for a moment that 
the turbulent viscosity hypothesis \(\mathcal{F} := -\nu_T(x)\frac{\partial}{\partial x}\avgF{u_y} \) holds~\cite{diamond91,pope01}. 
In this case Eq.~\eqref{eq:fullfmeanflowcombined} reduces to a simple advection-diffusion equation for ZFs
\begin{align}
\frac{\partial }{\partial t}  \avgF{u_{y}}  =   &
                        - \left(\avgF{u_{x}}+ V \right)\frac{\partial}{\partial x}  \avgF{ u_{y}}  
                        \nonumber \\&
                        +\frac{\partial}{\partial x} \left( \nu_T\frac{\partial}{\partial x}\avgF{u_y} \right)  
                        -  \avgF{u_{y}}  \frac{\avgR{  \Lambda }}{\avgR{  \ne }}
                        \nonumber \\&                                                    
                        +  \frac{\OmegaciN}{\avgR{  \ne }}\int_{x_0}^x d x \avgR{ \Lambda } ,
\end{align}
where the background density gradient pinch velocity \(V:=\nu_T/L_{\avgR{n}}\) appears now in addition to the radial outward pinch velocity \(\avgF{u_{x}}\). 
The direction of the additional pinch depends on the sign of the turbulent viscosity \(\nu_T\).

Finally, we extend the theory for energy transfer inside the kinetic \(\vec{E}\times \vec{B}\) energy \(E(t):= m_i \int dA\avgR{\ne\vec{u}_E^2}/2 \) to the full-F formalism. Here, the Favre decomposition \(E = E_0+ E_1\) is pivotal to derive the conservation laws for the zonal (or mean) 
\(E_0(t) :=   m_i \int dA\avgR{\ne}\avgF{u_y}^2/2 \)  and turbulent part \(E_1(t) :=   m_i \int dA\avgR{\ne}\flucF{\vec{u}}_E^2/2 \) 
of the kinetic \(\vec{E}\times \vec{B}\) energy  and supersedes the Reynolds decomposition in 
the \(\flucRrel{f}\) formalism~\cite{scott05,madsen17}. With the help of Eqs.~\eqref{eq:fullflognemeanflow} and ~\eqref{eq:fullfmeanflowcombined} we 
obtain the conservation laws for the zonal and turbulent  kinetic \(\vec{E}\times \vec{B}\) energy
\begin{subequations}
\begin{eqnarray}\label{eq:dtE0}
 \frac{\partial}{\partial t} E_0 =& \int dA \hspace{1 mm} 
 m_i\Big(
      \avgR{\ne} \mathcal{F}  \frac{\partial}{\partial x} \avgF{ u_{y}}
    - \frac{\avgF{u_{y}}^2}{2}  \avgR{  \Lambda } 
    \nonumber \\    &
    + \OmegaciN  \avgF{ u_{y}}\int_{x_0}^x d x \avgR{ \Lambda } 
 \Big) ,
 \\
\label{eq:dtE1}
\qquad
 \frac{\partial}{\partial t} E_1 =&
\int dA \hspace{1 mm} 
 m_i\Big(
 - \avgR{\ne} \mathcal{F}  \frac{\partial}{\partial x} \avgF{ u_{y}}
    + \frac{\avgF{u_{y}}^2}{2}  \avgR{  \Lambda } 
    \nonumber \\    &
    - \OmegaciN  \avgF{ u_{y}}\int_{x_0}^x d x \avgR{ \Lambda } 
     - \frac{e}{m_i}\avgR{\phi \Lambda}\Big) .
\end{eqnarray}
\end{subequations}
This unveils that the Favre stress term \( \avgR{\ne} \mathcal{F}  \frac{\partial}{\partial x} \avgF{ u_{y}}\) is the central  mechanism for energy transfer between the zonal and turbulent kinetic \(\vec{E}\times \vec{B}\) energy. As a consequence, density fluctuations (cf. Eq.~\eqref{eq:Favrestress}) manifest as 
 an additional transfer channel in the full-F formalism.
\section{Parallel closures}
Self-sustained drift wave turbulence  is maintained by the non-adiabatic parallel coupling of the relative density fluctuations and the electric potential, 
which can arise due to various mechanisms. 
Here, we exemplarily consider resistive drift wave turbulence, which arises due to resistive friction between electrons and ions along the magnetic field line.
This mechanism enters the 2D gyro-fluid models via the parallel closure terms (\(\Lambda_\delta \) or \(\Lambda\)) of the Hasegawa-Wakatani (HW) type as summarized in Table~\ref{table:hwmhwtable}.
\begin{table}[!ht]
 \begin{ruledtabular}
  \caption{HW closures for \(\flucRrel{f}\) and full-F models.}
 \label{table:hwmhwtable}
 \begin{tabular}{ l  l l }
   & ordinary HW &  modified HW \\
     \hline		
   \hspace{- 2mm} \(\Lambda_\delta/(\alpha_\delta \OmegaciN)\) {    }&  \hspace{- 2mm}  \( e\phi/\teN-\flucRrel{\ne} \)~\cite{hasegawa83,wakatani84,wakatani87} &   \hspace{- 2mm} \(e\flucR{\phi}/\teN- \flucR{ \flucRrel{\ne}}\)~\cite{numata07}\\
 \({\Lambda}/(\alpha \neref \OmegaciN)\) &   \hspace{- 2mm} \( e\phi/\teN -\ln\left(\ne/\avgR{  \ne }\right)  \)          &   \hspace{- 2mm} \(e\flucR{\phi}/\teN-\flucR{\ln\left(\ne\right)}\)  \\
\end{tabular}
\end{ruledtabular} 
\end{table}
Here, we introduced the full-F adiabaticity parameter \(\alpha:=  \teN  k_\parallel^2 /( \eta_\parallel e^2   \neref \OmegaciN )\)  with parallel wavenumber \(k_\parallel\)
and parallel Spitzer resistivity \(\eta_\parallel := 0.51 m_e \nu_e/(\ne e^2) \)~\cite{spitzer56,lingam17}. 
In the electron collision frequency \(\nu_e \) the Coulomb logarithm is treated as a constant 
so that \(\eta_\parallel\) has no explicit dependence on \(\ne\). 
As opposed to this in \(\flucRrel{f}\) models the density dependence in the collision frequency \(\nu_e (\ne) \approx \nu_{e0}\) is completely 
 neglected so that \(\alpha_\delta := \teN k_\parallel^2 /(0.51 m_e \nu_{e0} \OmegaciN) \) reduces to a parameter. 
Only then, the poloidal variations of the adiabaticity parameters vanish
(\(\flucR{\alpha_\delta} = \flucR{\alpha} = 0\)), and the full-F and \(\flucRrel{f}\) closures coincide in the limit of \(\avgR{  \ne } \approx n_G\) and \(\flucRrel{\ne} \ll 1\).
\section{Simulations}
We use the open source library \textsc{Feltor}~\cite{feltor40} to numerically solve the full-F gyro-fluid Eqs.
~\eqref{eq:ffmodel} with the modified HW parallel closure of Table~\ref{table:hwmhwtable}. 
Numerical stability is ensured by adding hyperdiffusive terms of second order \(- \nu \vec{\nabla}_\perp^4 n\) and  \(- \nu \vec{\nabla}_\perp^4 N\) to the right hand side of 
Eqs.~\eqref{eq:fullFdtne} and~\eqref{eq:fullFdtNi}.
Moreover, we append the right hand side of Eqs~\eqref{eq:fullFdtne} and~\eqref{eq:fullFdtNi} by a density source of the form 
\(\omega_S z\hspace{0.5mm} \Theta \left(z\right) \)  with \(z:=g(x) \left(n_G - \avgR{\ne}\right)\)
to maintain the initial profile in a small region \(x \in \left[0,x_b\right]\). Here, we defined the  Heaviside function \(\Theta(z)\) and  
\(g(x) := \left[1-\tanh{(x-x_b)/\sigma_b}\right]/2 \). The corresponding parameters are 
fixed to \(\nu=5\times 10^{-4} \csN \rhoN^3\), \(\omega_S = 0.1 \OmegaciN \), \(x_b= 0.1 L_x\) and \(\sigma_b = 0.5 \rhoN\) 
with cold ion sound speed \(\csN:=\rhoN\OmegaciN \).
The box with size \(L_x=L_y=128 \rhoN\) is resolved by a discontinuous Galerkin discretization with \(P=3\) polynomial coefficients and at least \(N_x=N_y=256\) equidistant grid cells.
The initial (gyro-center) density fields \(n(\vec{x},0)=N(\vec{x},0) = n_G(x)\left(1+\flucRrel{n}_0(\vec{x})\right)\) consist of the reference background density profile \(n_G\), 
which is perturbed by a turbulent bath \(\flucRrel{n}_0(\vec{x})\).

\begin{figure}[htb]
    \includegraphics[width= 0.485\textwidth]{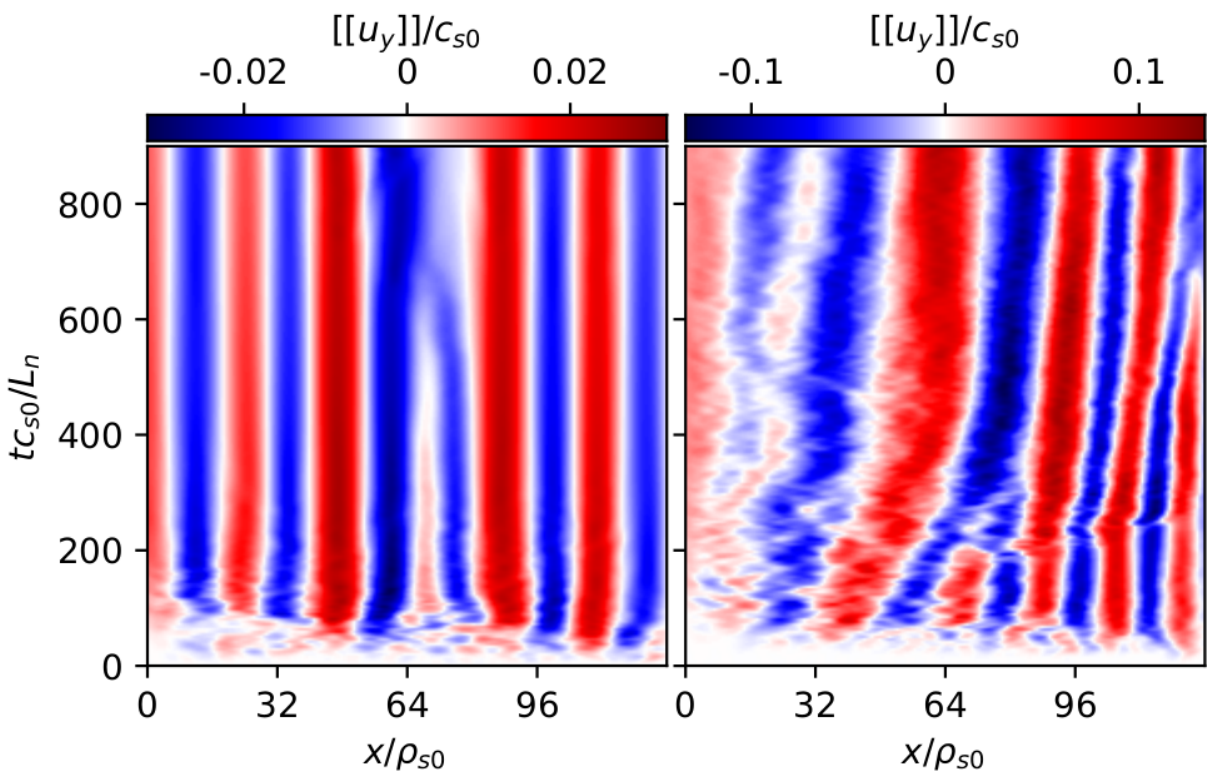}
    \caption{The spatio-temporal ZF evolution of the Favre averaged poloidal velocity \(\avgF{u_y}\) is shown for two different reference density gradient lengths 
    \(L_n =\left\{ 128,32\right\} \rhoN\) (left,right) in the high collisionality regime (\(\alpha=0.0005\)). Radial outward ZF advection occurs in the steep gradient regime (right).
    }
    \label{fig:radialpinch}
\end{figure}

NOB effects on drift wave-ZF dynamics, as it is described by Eqs.~\eqref{eq:fullflognemeanflow} and~\eqref{eq:fullfmeanflowcombined} with \(\langle \Lambda \rangle=0\), are 
in this setup studied by varying the adiabaticity parameter (or inverse collisionality) \(\alpha\) and the reference background gradient length \(L_n\).
In Fig.~\ref{fig:radialpinch} we show that \(L_n\) crucially determines the time evolution of ZFs 
in the high collisionality regime with \(\alpha=0.0005\). While stationary ZFs emerge for \(L_n=128  \rhoN\), a radial outward pinch of ZFs occurs 
for a four times smaller reference background density gradient length \(L_n=32 \rhoN\).

\begin{figure}[htpb]
\centering
\includegraphics[width= 0.485\textwidth]{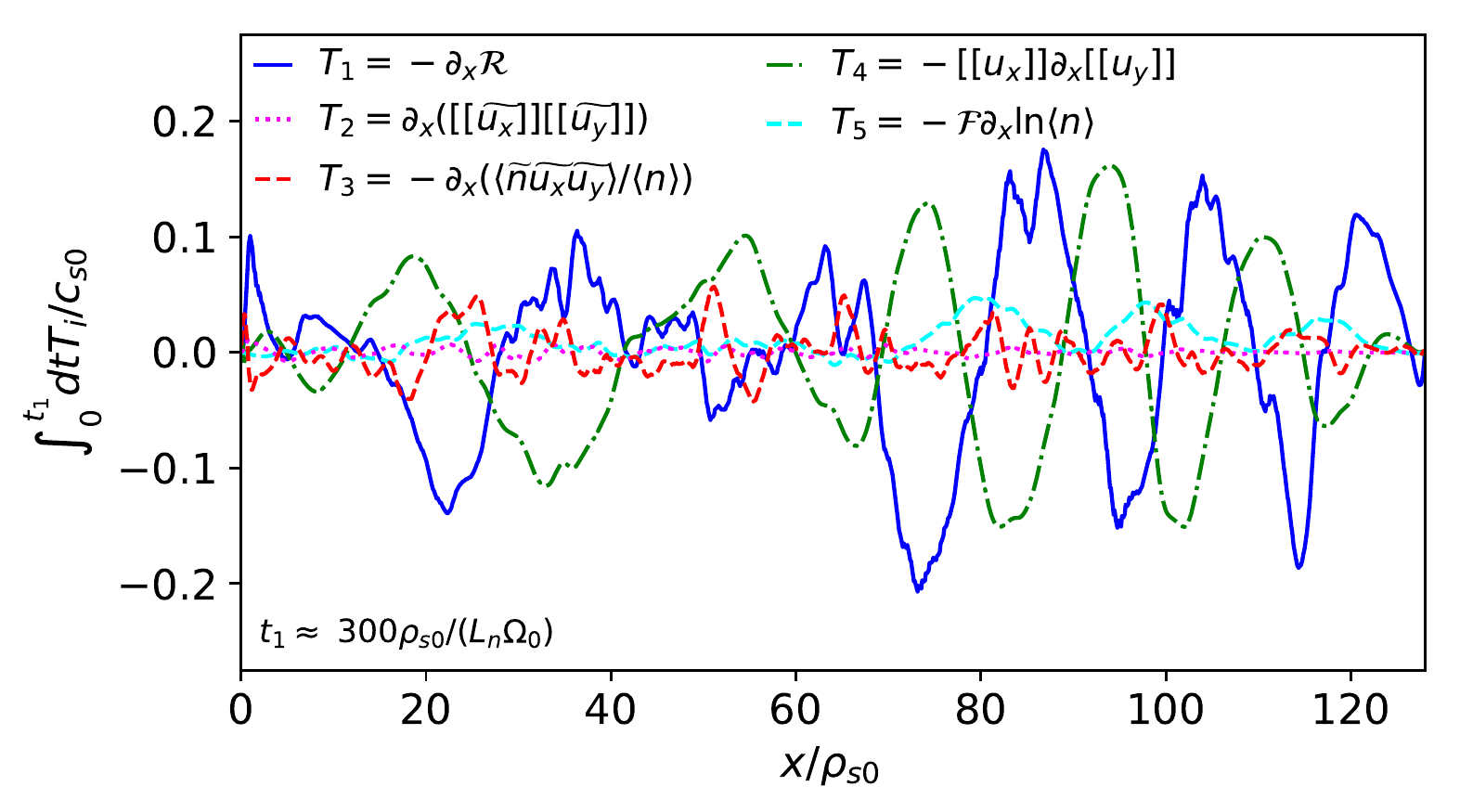}
\caption{The radial profile of the terms of the right hand side of Eq.~\eqref{eq:fullfmeanflowcombined}  for \(\alpha=0.0005\) and \(L_n=32 \rhoN\). 
The ZF signature of the radial advection term \(T_4\) is comparable to the Reynolds stress \(T_1\).
}
\label{fig:stressesvsx}
\end{figure}

In this steep gradient and high collisionality regime the ZF signature is no longer solely determined by the conventional Reynolds stress drive, 
which is illustrated in Fig.~\ref{fig:stressesvsx}. The Reynolds stress drive \(T_1\) is here comparable to the radial advection term  
\(T_4\), which explains the observed radial outward propagation of ZFs in Fig.~\ref{fig:radialpinch}.

In the following the parametric dependence of each term \(T_i\) on the right hand side of Eq.~\eqref{eq:fullfmeanflowcombined} is investigated. 
To this end, the contribution of each term \(T_i\) on ZF evolution is measured by taking the \(L^2\) norm, denoted by \(\big\| h\big\|_{2}\), 
 of the time integrated contribution.
Following this, we propose a measure of the relative ZF contribution
\begin{eqnarray}
 M_i&:=\frac{\big\| \int_0^{t_1} dt \hspace{0.5mm}  T_i\big\|_{2} }{\sum_{j=1}^{5}\big\| \int_0^{t_1} dt \hspace{0.5mm} T_j\big\|_{2}}.
\end{eqnarray}

\begin{figure}[htpb]
\centering
\includegraphics[width= 0.485\textwidth]{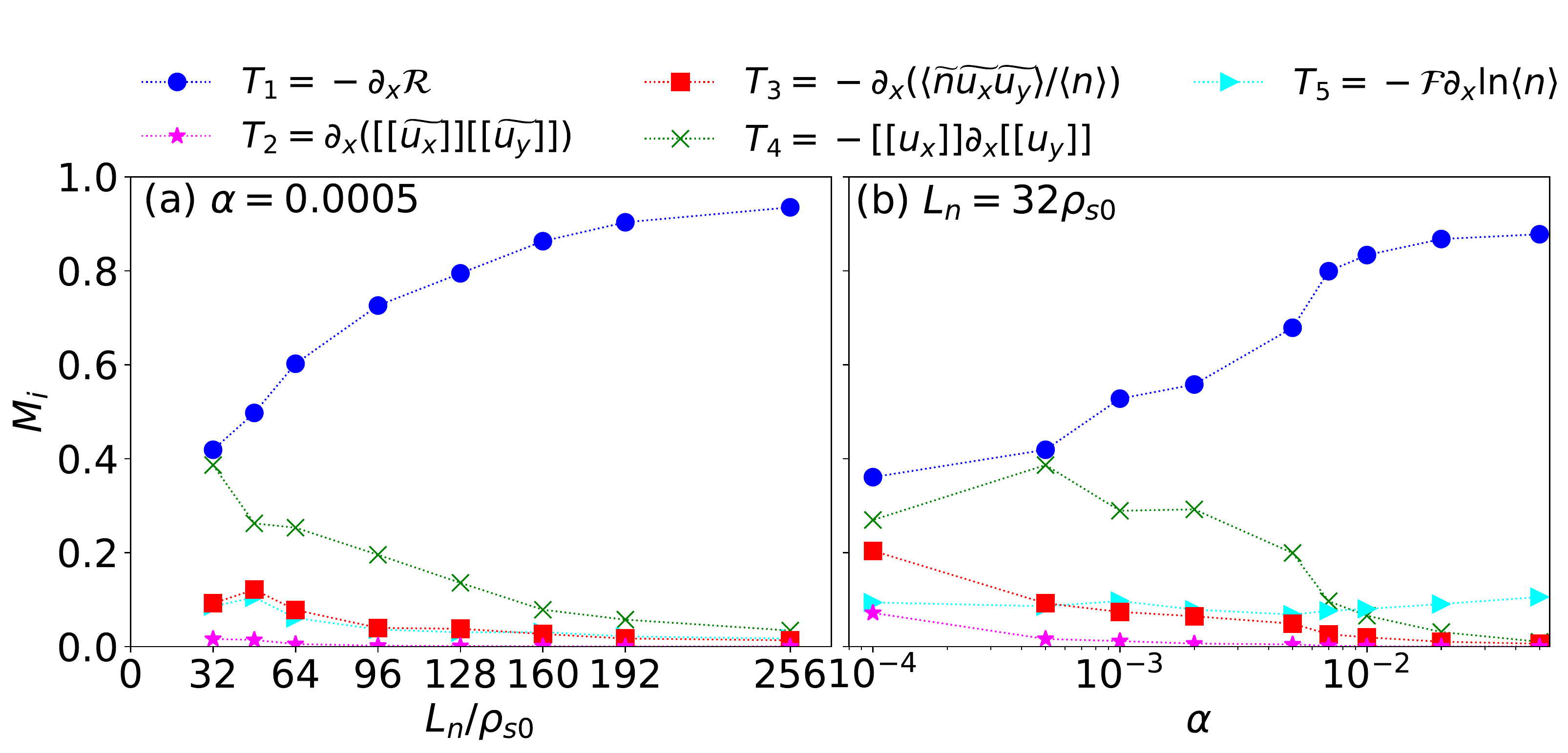}
\caption{(a) The NOB ZF terms decrease with the reference background gradient length \(L_n\) in the high collisionality regime (\(\alpha=0.0005\)). 
(b) For a fixed \(L_n=32 \rhoN\) all the NOB ZF terms significantly contribute to the ZF dynamics 
in the high collisionality regime. As opposed to this, only the background density gradient drive \(T_5\) remains alongside the Reynolds stress drive \(T_1\)
in the small collisionality regime.
  }
\label{fig:termsvsLnalpha}
\end{figure}

In Fig.~\ref{fig:termsvsLnalpha}a we show that the relative contribution \(M_i\) of the NOB ZF terms (\(T_2,\dots,T_5\)) 
decreases with the reference background density gradient length \(L_n\) in the high collisionality regime (\(\alpha=0.0005\)). 
The summed up relative contribution of the NOB ZF terms exceeds the one of the conventional Reynolds stress for  \(L_n=32 \rhoN\).
For steep reference background density gradients the radial advection term  \(T_4\)
exhibits the largest relative contribution to the ZF dynamics of all the NOB terms. 

In Fig.~\ref{fig:termsvsLnalpha}b the dependence of the relative importance of each term on the adiabaticity parameter \(\alpha\)
is depicted for a fixed reference background density gradient length \(L_n=32\rhoN\). 
While the conventional Reynolds stress term is again the dominating ZF contributor in particular for small collisionalities, 
all the NOB terms, except the background density gradient drive  \(T_5\), gain in importance for higher collisionalities. 
Interestingly, in the small collisionality regime (\(\alpha\geq0.01\)) the background density gradient drive \(T_5\) exceeds all the remaining NOB actors. 
The quadruple fluctuation drive \(T_2 \) is for all studied parameters the smallest contributor to the ZF dynamics.

\begin{figure}[htpb]
    \centering
  \includegraphics[width= 0.485\textwidth]{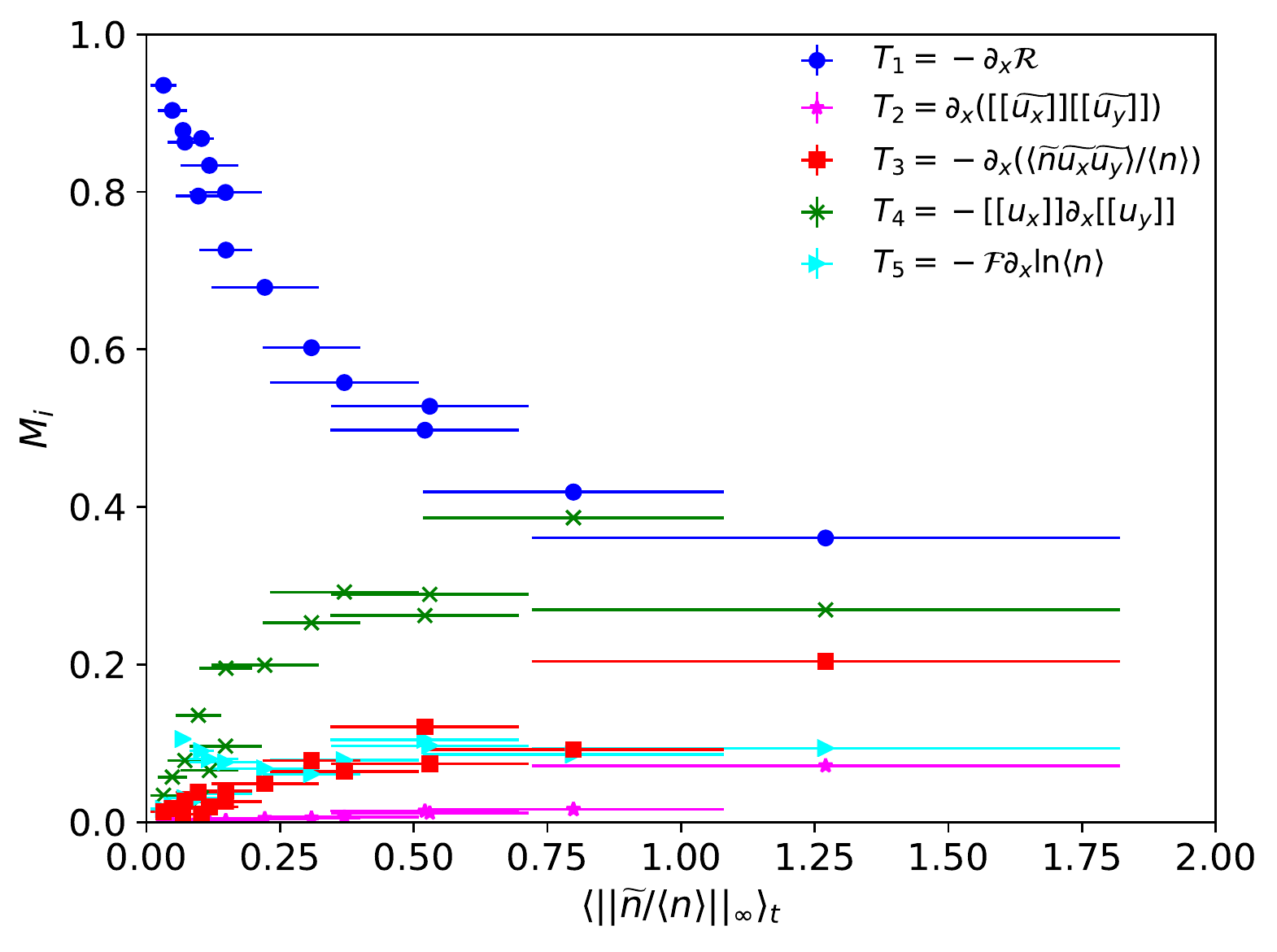}
    \caption{The relative contributions of the NOB ZF terms increase with the relative density fluctuation amplitude. 
    In particular they can amount to roughly two thirds of the ZF dynamics.  }    
    \label{fig:termsvsflucandgamma}
\end{figure}

The dependence of the ZF terms on the time averaged maximum of the relative density fluctuation level 
\(\langle \big\| \flucR{\ne}/\avgR{\ne} \big\|_\infty \rangle_t \) is shown in Fig.~\ref{fig:termsvsflucandgamma}.
Here, we denote the time average by  \(\langle h \rangle_t \) and compute the maximum with the help of the supremum norm  \(\big\| h \big\|_\infty\).
In Fig.~\ref{fig:termsvsflucandgamma} the conventional Reynolds stress drive \(T_1\) contribution weakens with increasing relative density fluctuation level. The radial advection term  
\(T_4\) and the triple fluctuation drive 
\(T_3\) are the dominating NOB ZF contributors for high relative density fluctuations,
while the background density gradient drive \(T_5\) can be relevant likewise for small relative density fluctuations.
\section{Conclusion}
We have generalized the ZF equation~\eqref{eq:deltafmeanflow} to account for NOB effects in Eq.~\eqref{eq:fullfmeanflowcombined}. 
Most importantly, the former Reynolds stress \(\mathcal{R}\) is replaced by the Favre stress \( \mathcal{F} \), which adds to its predecessor in case of high relative density fluctuations. 
The latter is accompanied by two new agents in the NOB ZF Eq.~\eqref{eq:fullfmeanflowcombined}.
The first of these radially advects ZFs by the Favre averaged radial drift velocity, which is proportional to the averaged radial particle flux. 
The second term scales inversely with the background density gradient length and affects the ZF dynamics even if the relative density fluctuations are small or if the
Favre stress is radially homogeneous. 
Thus, this term may be of significance in or 
during the formation of radial transport barriers, where steep density profiles form with strongly reduced radial particle transport.

Additionally we extended the ordinary and modified HW model to the full-F theory. 
We simulated the full-F gyro-fluid model with the modified HW closure to  numerically corroborate our theoretical results.
The simulations successfully reproduced the predicted radial advection of ZFs, which appeared for small reference background density gradient lengths 
and large averaged radial particle flux.
Moreover, our numerical parameter study showed that the NOB ZF drives can be comparable to the Reynolds stress drive in the herein scanned parameter range.
In particular the deviation between the Reynolds and Favre stress drive increases with the relative density 
fluctuation amplitude, collisionality and inversely with the reference background density gradient length. 
This deviation is mainly reasoned in  the triple fluctuation drive. 
Its importance in steep background density gradient regimes is in qualitative agreement 
with the theoretical estimate in the strong turbulence regime~\cite{wang16}.
A similar dependence as for the Favre stress drive is found for the radial ZF advection mechanism.
For the background density gradient drive only a dependence on the reference background density gradient is observed.

The presented results strongly argue in favor of the development and application of
full-F gyro-fluid or gyro-kinetic models for simulation of fusion edge plasma
turbulence, and in general demonstrate exemplarily the relevance of NOB
effects for ZF formation in fluids and plasmas with large
fluctuations and inhomogeneities.
The latter conditions prevail e.g. during the low- to high-confinement mode transition.
Thus, a consistent full-F simulation approach of this phenomenon is crucial to allow for the herein presented NOB ZF mechanisms.

Finally, we emphasize that the relative error between the Favre and Reynolds average of the poloidal velocity, derived to \(|\avgF{\flucR{u_y}}/\avgR{u_y}|\), is typically below a few percent. Thus, our proposed NOB ZF theory is also applicable to experimental measurements of the Reynolds averaged poloidal velocity \(\avgR{u_y}\).
\section{Acknowledgements}
This work was supported by the Austrian Science Fund (FWF) Y398. 
R. K. was supported with financial subvention from the Research Council of Norway under grant
240510/F20.  The computational results presented have been achieved using the Vienna Scientific Cluster (VSC) and 
the EUROfusion High Performance Computer (Marconi-Fusion).
This work has been carried out within the framework of the EUROfusion Consortium and has received funding 
from the Euratom research and training programme 2014–2018 under grant agreement No 633053. 
The views and opinions expressed herein do not necessarily reflect those of the European Commission. 
\bibliography{meanflows_aip.bib}
\bibliographystyle{aipnum4-1.bst}

\end{document}